\documentclass[12pt]{article}
\newcommand{\mathsym}[1]{{}}
\def\id{\protect{{1 \kern-.28em {\rm l}}}}

\def\be{\begin{eqnarray}}
\def\ee{\end{eqnarray}}

\makeatletter
\renewcommand\section{\@startsection {section}{1}{\z@}%
                                   {-3.5ex \@plus -1ex \@minus -.2ex}%
                                   {2.3ex \@plus.2ex}%
                                   {\normalfont\large\bfseries}}
\renewcommand\subsection{\@startsection{subsection}{2}{\z@}%
                                   {-3.25ex\@plus -1ex \@minus -.2ex}%
                                   {1.5ex \@plus .2ex}%
                                   {\normalfont\normalsize\bfseries}}

\makeatother

\def\Tr{{\rm Tr}}

\def \foot {\footnote}
\def \bi{\bibitem}

\def \ha {{1 \over 2}}

\def \ci{\cite}
\def \N {{\mathcal N}}

\def\S{{\mathcal S} }

\def \E {{\mathcal  E}} \def \J {{\mathcal  J}}

\def\C{{\bf C}}

\def\g{\gamma}
\def\s{\sigma}

\def\ov{\over}

\def\J{{\mathcal J}}

\def\E{{\mathcal E}}

\def\l{\lambda}

\def \k {\kappa}
\def\foot{\footnote}

\def \ci {\cite}

\def \foot {\footnote}
\def \bi{\bibitem}
\def \ha {{1 \over 2}}

\def \fo { {1\ov 4}}

\def \Tr {{\rm Tr}}

\def \l  {\lambda}

\def \N {{\mathcal N}}
\def \S {{\rm S}}

\def \D {\Delta}
\def \N {{\mathcal N}}

\def \bi{\bibitem}
\def \la {\label}

\def \l {\lambda}
\def\foot{\footnote}

\def \sql {{\sqrt \l}}
\def \adss {$AdS_5 \times S^5~$ }
\newcommand{\rf}[1]{(\ref{#1})}
\def \ov {\over}

\def\N{{\cal N}}

\def\cc{\circ}

\def \ha{{1\ov 2}}

\def \J {\mathcal{J}}

\def \E {{\cal E}}

\def \S {{\cal S}}
\def \J {{\cal J}}

 \def \bb {\bar \beta}

\def \bi{\bibitem}
\def \la {\label}

\def \l {\lambda}
\def\foot{\footnote}

\def \sql {{\sqrt \l}}

\def \adss {$AdS_5 \times S^5$\ }

\def \D {\Delta}

\def \ov {\over}

\def \varpi {{\rm w}}
\def \OO {{\cal O}}

\def \te {\theta}

\def \cc {{\rm f}}

\def \S  {{\rm S}}

\def \C {{\cal C}}

\def\Tr{{\rm Tr}}

\def \s {\sigma}

 \def \J {{\cal J}}
 \def \S {{\cal S}}
 \def \E {{\cal E}}

 \def \sql {\sqrt{\lambda}}

\begin{document}


\overfullrule=0pt
\parskip=2pt
\parindent=12pt
\headheight=0in \headsep=0in \topmargin=0in \oddsidemargin=0in

\vspace{ -3cm}
\thispagestyle{empty}
\vspace{-1cm}

\rightline{Imperial-TP-AT-2011-1}


\begin{center}
\vspace{1cm}
{\Large\bf  

Semiclassical string computation  \\
 of  strong-coupling corrections  to  
dimensions   of operators in  Konishi multiplet

\vspace{1.2cm}

   }

\vspace{.2cm}
 {
R. Roiban$^{a,}$\footnote{radu@phys.psu.edu} and
 A.A. Tseytlin$^{b,}$\footnote{Also at Lebedev  Institute, Moscow. tseytlin@imperial.ac.uk }}\\

\vskip 0.6cm

{\em 
$^{a}$Department of Physics, The Pennsylvania  State University,\\
University Park, PA 16802 , USA\\
\vskip 0.08cm
\vskip 0.08cm $^{b}$Blackett Laboratory, Imperial College,
London SW7 2AZ, U.K.
 }

\vspace{.2cm}

\end{center}

\begin{abstract}
 
 Following our earlier work in arXiv:0906.4294 we show how to use semiclassical string quantization approach to compute the leading corrections to the energy of 
 $AdS_5\times S^5$ string states on the first excited string level that should correspond to operators in the Konishi multiplet of $\N=4$ SYM theory. Compared to examples in our previous paper the string solutions we consider here carry an extra component of $S^5$ angular momentum $J$. This facilitates their identification with operators in the Konishi multiplet. We show that for such string states with $J=2$ the coefficient of the subleading 
 $\lambda^{-1/4}$ term in the large string tension expansion of the energy is twice the one found in arXiv:0906.4294. The resulting value matches the one found for the Konishi state in the $sl(2)$ sector from the Y-system/TBA approach, resolving an apparent disagreement claimed earlier.

\end{abstract}

\newpage
\setcounter{equation}{0} 
\setcounter{footnote}{0}
\setcounter{section}{0}

\renewcommand{\theequation}{1.\arabic{equation}}
 \setcounter{equation}{0}

\setcounter{equation}{0} \setcounter{footnote}{0}
\setcounter{section}{0}

\def \edd {\end{document}}

\def \cc {{c }} 
\def \OO {{\cal O}}
\def \te {\textstyle}
\def \fl {\sqrt[4]{\l}}

\def \fo {{\textstyle{1 \ov4}}}
\def \rx {{\rm x}}
\def \hg {{\hat \g}}

\def \C  {{\rm C}}
\def \hC  {{\rm \hat  C}}
\def \dd  {{\rm d}}
\def \bb {{\rm b}}
\def \dDelta {2}
\def \sql {{\sqrt{\l}}}

 \def \an {{\rm an}} \def \nan {{\rm nan}}
 

\section{Introduction }

This  paper is a direct  sequel to our earlier   work \ci{rt} where  we  discussed   how a 
semiclassical quantization  of ``short'' strings in \adss  may be used to determine 
 the structure of large string tension ($T= { \sql \ov 2\pi} \gg 1$)
  expansion of the AdS energy  of 
quantum string states on the first excited string level. The  analysis 
of the structure of the  ``short'' string expansion  \ci{tir,rt} or of
 the marginality condition for string
vertex operators \ci{rt,tss} implies   that the energy of such 
states   for  large $\sql$  may be written as 
\be \la{fow}
E
=  2 \fl   \  + b_0  +  {b_1 \ov \fl } + {b_2 \ov \sql  } +  \OO( {1 \ov (\fl)^3  } )  \ .   \ee
The  coefficients $ b_1, b_2, ...$  should  be  the same  
 for all  states in one supermultiplet 
with $b_0$ being an integer related to a position of a given state 
in the supermultiplet. As was argued in 
\ci{rt}, for the Konishi multiplet $b_0 +4 = \D_0$   should be equal to 
bare ($\l=0$) 
dimension of the corresponding gauge-theory operator,   while  $b_1$  should receive  contributions
only from  the classical string  energy  and the string 1-loop    correction to it.

Generic  string  solutions  \ci{gkp,ft1,ft2}  carry  global 
  $SO(2,4) \times SO(6)$  charges  $ (E, S_1, S_2; J_1, J_2, J_3)$, i.e. the 
  energy and  components  of spins/orbital momenta  in 2+3 orthogonal planes 
  of  $AdS_5\times S^5$.
  While in the standard semiclassical expansion the values  of these charges 
  are assumed to be of order of  the string tension which is  taken to be large, one may 
  formally  interpolate the  semiclassical results to fixed values of $S_r$ and $J_i$ 
  by considering   ``short'' string limit \ci{tir,rt}  when $ \S_r = {S_r\ov \sql}\ll 1 $
  and $ \J_i = {J_i\ov \sql}\ll  1$. 
  
Establishing a   correspondence      between  semiclassical string states   and the  dual 
 gauge-theory operators is, in general, a  highly non-trivial  enterprise. 
 The  values of the global  charges   may be used as a guide in constructing such  a 
 map.
 In particular, 
 quantum string states and the dual gauge theory operators
  should be highest-weight states with Dynkin labels\foot{We follow 
 the standard  notation used in \ci{rt}: 
   $SU(2)\times SU(2)$ labels $(s_L,s_R)$ for $SO(4)$ 
and the  Dynkin labels $[p_1,q,p_2]$ for $SU(4)$ are given by 
$s_{L,R} = \ha (S_1\pm S_2),$ and 
$p_{1,2}= J_2\mp J_3, \ q = J_1-J_2$.}

\noindent
$ [ J_2-J_3,  J_1-J_2 , J_2+J_3]_{( {S_1+S_2\ov 2}, {S_1-S_2\ov 2})} $.
 For example, the singlet  operator $\Tr ( \bar\Phi^i \Phi_i )$ which is the 
 ``ground-state'' of the Konishi multiplet \ci{ferr,bia}\foot{We
  copy the  
  table of its  states (in the form taken from \ci{bia}) at the end of this paper.}
 corresponds to   $[0,0,0]_{(0,0)}$  with $\D_0=2$  while 
 its supersymmetry descendants in the $sl(2)$ and $su(2)$ sectors are, respectively, 
 $\Tr (\Phi_1  D^2 \Phi_1)$   with $S=J_1=2$  or   $[0,2,0]_{(1,1)}$, 
 and $\Tr ([\Phi_1,\Phi_2]^2)$  with $J_1=J_2=2$  or   $[2,0,2]_{(0,0)}$,  both with
 bare  dimension  $\D_0=4$. 
 
  The semiclassical description of  string states is very crude (unless the values of 
  all hidden charges are  specified, as it is the case in the integrability-based 
   description). 
  In particular,  one may not be able to distinguish between zero and finite 
  values of spins at large string  tension,  so that 
  it is important to  have additional cross-checks. 
  In particular, one should  verify 
  that the string theory predicts 
   universal (i.e. equal) 
    values of anomalous dimensions for states that 
  are expected to belong to the same supermultiplet.
  Such a universal behavior was observed  in \ci{rt}. The states 
  considered there  formally fit (under some particular  assignment
   of their global charges)
  into  some representations present in the Konishi multiplet table.
  There was, however,  
  only a  circumstantial  
  evidence\foot{For example, one may check that 
  in the flat space limit the small circular string states considered in \ci{rt} 
   apparently belong to the 
  same supersymmetry multiplet.}  that these states  correspond indeed to  members 
  of the Konishi multiplet,
  implying that one needs  some additional input  or data. 
   %
  %
  
  An additional assumption  used in \ci{rt} was that in identifying 
  semiclassical spinning string states   with quantum string states  one  may 
  use analogy with  the  bosonic or  NSR  string  spectrum in flat space 
  $E^2 = 2 \sql (N-2)$ where $N=2,4,...$ is the maximal value of spin at a given string level 
(i.e. $N=2$  at the massless level, $N=4$ at the first excited string level, etc.).
Following such  pattern would  suggest that one should shift the value of the total 
spin of  a semiclassical string by -2 when comparing to  quantum string states. 
However, in describing  the light-cone GS superstring spectrum  one 
 identifies  the  massless  string  level with the ground  state 
not carrying any genuine string  oscillators: all  massless (supergravity) string  modes
  are obtained by quantizing the superparticle representing 
  the point-like string limit  (with 
  the  spin of the graviton  carried  by the fermion 0-modes building 
 the vacuum supermultiplet, etc.; the same picture applies 
 in the case of the   superstring in a pp-wave background \ci{bmn,mts}).
   Then   the 
     spin of an  extended string solution   should be associated 
only with ``extra'' oscillators   acting on the ground state.

In this  paper we shall  use   this  superstring pattern when 
comparing   semiclassical  strings to quantum string states, i.e. we shall  assume that 
in the flat-space limit 
$E^2 = 2 \sql N$,  \   $N=0,2, ...$, 
   with the  maximal spin 
carried  by extended string motion  being $N$.
   States on the first excited string level $N=2$
   will have just two oscillators, and the resulting values of spins they may carry
   are
  $S_1=2,\ S_2=J_i=0$,  or $S_1=S_2=1,\ J_i=0$, or  
  $S_1=S_2=0, \ J_1=2$,  or $S_1=J_1=1,\ S_2=J_i=0$, or $J_1=J_2=1,\ S_i=J_3=0$. 
 States  with such small values of  spins will not correspond to  non-trivial states 
 in the  ``minimal'' closed  $sl(2)$ or  $su(2)$ sectors on the gauge  theory side. 
 Also, identification to  members of  the Konishi multiplet table  appears to be 
 ambiguous. 
 
 There is, however,  a  natural resolution: instead of considering classical solutions
  that carry no orbital angular momentum $J=J_3$ in $S^5$ 
 (as was  the case   in \ci{rt}) one  should   switch on a minimal non-zero
 value of $J$. While having    $J\not=0$ will not influence the  value of the string level 
 it will allow for  a natural identification of the corresponding semiclassical string solutions 
 with  operators in  the  $sl(2)$ or  $su(2)$ sectors and with 
 members in the Konishi multiplet (for $J=2$). 
 
Thus the   main  point of the present paper  is  that one  should 
   focus on the simplest extended string  states built ``on top'' of the BPS 
   vacuum state with charge $J=2$
  (i.e. the chiral primary  operator $\Tr (\Phi_1^2)$ on the gauge theory side 
  or a superparticle state 
  on the string side).
  For this reason, to describe 
  string   states dual to the Konishi multiplet states  
  here    we shall  consider  simplest  
  string  solutions 
   that  carry a (minimal)   non-zero value of the $S^5$  orbital momentum,  
   i.e. their  center of mass will  be  orbiting a circle in $S^5$.\foot{As we shall
   discuss below, the simplest  ``short'' strings without 
  such $S^5$ momentum  like folded spinning strings or pulsating strings with just
  one quantum number  (for which $b_1$ in \rf{fow} happens to be 
  transcendental \ci{tir,bef,bec}) presumably do  not to  correspond to members of the 
  Konishi multiplet.}

  As we shall see below,  the direct generalizations 
  of the semiclassical string states discussed in \ci{rt} to states that 
  carry  an   additional momentum $J=2 \ll \sql$ have their natural places
   in the Konishi 
  multiplet table of 
  states.\foot{The state with $J=2$ in the supergravity multiplet, 
  dual to a chiral primary  operator, may be  described  (like the graviton 
  in the superstring spectrum)  as a state created 
  by four fermion zero modes.  
At the same time, in our 
 semiclassical description,  the $S^5$  momentum $J$ is captured by the orbital
 motion of the string's center of mass. 
 It is possible, however,  to choose a different quantization scheme
  \cite{mts} in which the 
 ``bottom" of the supergravity multiplet is the $E=J=2$ state 
 that  carries no fermion  zero modes and thus admits a 
   semiclassical description  in terms  of a massless particle 
   orbiting a big circle of $S^5$. Then excited string states 
 obtained by acting with  bosonic creation operators on this ``vacuum''  
 should  also  have a semiclassical description. 
 The same should hold both for the ``bottom" of the  excited string multiplet as well as
  for states at higher levels. Indeed, by acting with 
 supersymmetry transformations on the lowest state in the multiplet 
 one always obtains at  least one term in which all supersymmetry transformations act on 
 the bosonic oscillators  and which  thus does not contain  fermion zero modes
 (a similar observation applies to the type IIB superparticle multiplet \ci{rrm}). 
 For an even number of such transformations there will be at
  least one component which is  bosonic (i.e. it does not
   contain an even number of fermion oscillators) which  
   should thus have a semiclassical counterpart.} 
  The corresponding  value of $b_1$  in \rf{fow}  is 
  shifted  by the ``classical'' contribution proportional to $J^2$ as 
  \be 
  b_1(J) = b_1(0)  + {1 \ov 4} J^2  \ . 
  \la{bb}
  \ee 
  As the  universal  value of the  coefficient  $b_1(0)$ found in \ci{rt} 
  for  different solutions was 1, this implies then that  $b_1(2)=2$. 
  This  is precisely  the value   for the 
  Konishi multiplet  state in the $sl(2)$
  sector (having  $S=J=2$) found  in \ci{gkv}, and recently also in  \ci{fro}, 
   from the  
  Y-system/TBA equations   (starting from 
  the weak coupling region and interpolating numerically to strong coupling).  
   
  The $J^2$ term in \rf{bb}   has a simple classical origin.\foot{This  term was 
    already mentioned   
  in footnote 34 in \ci{rt}. There we considered the example of a  short  $ (S,J)$  
  folded string 
  \ci{ft1,bet}  
    but we were wrongly  assuming that $S=4$   (instead of $S=2$)
       in  which case  there is no corresponding state  in the Konishi    multiplet.
   Similar  $J^2$ term  in $b_1$ was mentioned also in \ci{frol,afss}.}
    Consider, e.g.,  a ``short'' string 
   with  some charge $N$ (e.g. spin or oscillation number)  orbiting also in $S^5$. Then  
   for small $ {N \ov \sql}$  we have 
   for the  classical string   energy $E^2_{0} = 2 \sql N  +  a  N^2  + J^2 + ... $
   ($a$ is a state-dependent constant). 
    Assuming that $Q,  J \ll \sql$    we then find the following expansion  
   \be \la{rty}
   E_{0} = \sqrt{ 2\sql  N}\  \Big[ 1 +  { 1 \ov 4 \sql} \Big( a N  + { J^2 \ov N} \Big) 
    + ... \Big] \ .  \ee 
  For  $N=2$ we then  get a state on the first excited string level.  Switching on 
  non-zero $J$  thus shifts the  value  of 
    $b_1$  coefficient  in \rf{fow}  by the  $J^2$ term  given in \rf{bb}.
     
 Below we shall illustrate and verify this argument  on several explicit 
 examples. We shall consider the same  solutions as in   \ci{rt} and generalize   them to 
    the case of  non-zero  orbital momentum  $J$ in $S^5$. 
  In particular, 
    we will show  that for rigid circular strings  
     switching on a non-zero  value of  $J \ll \sql $  does not
    alter  the value of the 1-loop string correction to $b_1$ found in  \ci{rt}.
    The resulting states with  $N=2$ (to be on the first excited string level, cf. \rf{rty}) 
     and $J=2$  will   transform in the same representations as the 
      corresponding members of the 
    Konishi multiplet     and  for all of them we will find that  $b_1=2$,  
    matching   the Y-system  result of \ci{gkv}.  
%

\renewcommand{\theequation}{2.\arabic{equation}}
 \setcounter{equation}{0}

\section{Examples of semiclassical string states
dual to\\  $\D_0=6$ states in Konishi multiplet } 
 
 We shall start  with  examples of simple circular string 
 solutions  which have three non-vanishing spins and generalize 
 the 2-spin solutions considered  in  \ci{rt}.
%
 For each such solution (with values of spins
 representing a state on the first excited string
 level) there  will  be  a  state in the corresponding 
 representation in the  Konishi multiplet table.
%
 Members of the same supermultiplet are expected to have the same
   anomalous dimension 
 and we  will   see the 
 evidence of that  in the universality of the predicted value of 
 $b_1=2$ in \rf{fow}.  
 

\subsection{Small circular spinning string with $J_1=J_2$  and $J_3 \not=0$ }\label{J1J2J3}

  The ``small'' circular spinning string with $J_1=J_2=J'$   in $S^5$
  has  its classical energy given  simply 
  by its  flat-space expression \ci{rt}, i.e. 
  $E_{0} = \sqrt{2 \sql (J_1 + J_2) }=  2 \sqrt{ \sql J' } $. 
  Its  generalization to include a non-zero 
   orbital momentum $J=J_3$ in a 
  third plane is given by \ci{ft2,art}\foot{We follow closely the 
   notation of \ci{rt}.} 
  \be
&& X_1=\ a\  e^{i (w \tau + \s)}
~,~~~~~
X_2=\ a\  e^{i (w \tau  - \s)}
~,~~~~~
X_3= \sqrt{ 1 - 2 a^2}\    e^{i \nu  \tau }\ , \la{h} \\ 
&& 
{\cal E}_0^2 = \kappa^2 = 4 a^2 +  \nu^2={\nu^2+   \frac{4 \J'}{\sqrt{1+ \nu^2}}   } \ , 
~~~~~~~~~
w^2 = 1 + \nu^2\ , \la{hh} \\
&& 
\J'\equiv  \J_1=\J_2= a^2 w
~,  ~~~~~~
\J\equiv  \J_3= 
(1 - 2 a^2)\,\nu \ , \ \ \ \ \ 
\nu=  \frac{\J}{ 1 - { 2\J'\ov  \sqrt{1+ \nu^2} } } \ . \la{hhh}
\ee
Expanding the classical string energy for $ \J' = { J' \ov \sql} \ll 1, \ 
\J = { J \ov \sql} \ll 1$ we get   (cf. \rf{rty}) 
  \be
E_0=2\sqrt{\sql J'}\ 
\Big[1+\frac{1}{\sql}\frac{J^2}{8J'}-
\frac{1}{(\sql)^2}\Big(\frac{J^4}{128J'{}^2}-\frac{J^2}{4}\Big)+\dots\Big]  \ . \la{ec} 
\ee 
One can give a general argument (see sec.~\ref{argument} below) 
that 
the leading term in the  1-loop  correction to string energy
expanded in  $\J = { J \ov \sql} \ll 1$
should not depend   on $J$, i.e. should  have the same value as in the $J=0$ 
case considered in \ci{rt}.  For this particular solution 
we  explicitly  prove this in the Appendix (using fluctuation frequencies found in \ci{ft3}).
As a result, the  1-loop  corrected expression for the energy can be written as 
 \be 
 E=E_0+E_1  =  2 \sqrt{\sqrt{\lambda} J'}\ \Big[1+ 
  { 1 \ov \sql} \Big({ J^2 \ov 8  J' } + \frac{1}{ 2} \Big) 
 + \OO( { 1\ov (\sql)^2  }) \Big]  +  b_0 
    \ .   \la{py} \ee
    The value of the constant term $b_0$ is sensitive to  the 0-mode 
      contributions  and 
   appears to be discontinuous with $J$: equal to 0 for $J=0$ 
     but equal to 2  for $J=2$ 
      (cf. \ci{rt}).\foot{Here  we are assuming that 
      $J= {\cal O}(1) \ll \sql$; for large   $J\sim \sql $
      a linear  term in $J$   will appear  already in the classical string energy.
      In general, to systematically reproduce   such contributions to $b_0$ 
      one would need to perform   0-mode  superparticle quantization.}
              
   To get  a state  on the first excited string level we 
    should  choose $J'=1$, i.e.  $ J_1=J_2 =1$.\footnote{As already discussed in the 
    introduction, this  identification of 
    charges is different from the 
  one in \ci{rt} where it was assumed  that  $N=2J'$ should be  renormalized as $2J' 
  \to 2J'_{eff}= 2J'-2$.
   In general, fixing constant quantum integer  shifts  of charges (like energy and spins)
   is a subtle issue in the semiclassical  approach. Its  resolution requires to start  
   with the
   correct definition of the charge operators, see also \ci{rt}. The identification 
   in \ci{rt} was 
   based   on an NSR-type  interpretation of states, where the first 
   excited string level has two 
   left-moving and two right-moving oscillators. 
   %
   The first excited level of the GS string has only one  left-moving   and 
   one right-moving oscillator, which is what we are assuming  here.
   }
    Choosing     the minimal  non-trivial value of $J=J_3=2$ and 
    relabeling the  momenta  (as $J_1=2, J_2=J_3=1$) to compare to states 
    in the Konishi multiplet table, we conclude   
 that, remarkably,  there is a
  unique corresponding state there --  $[0,1,2]_{(0,0)}$
 (at level $\D_0=6$).
  The resulting  value of $b_1$ in \rf{fow}  that follows from \rf{py}
  is then consistent with \rf{bb}, 
  \be
  b_1 = 2 \Big( { J^2 \ov 8  J' } + \frac{1}{ 2}\Big)_{{J=2, J'=1}}  =2 \ .\la{v} \ee

\subsection{ Small circular spinning  string with $S_1=S_2$  and $J\not=0$  }

The same discussion can be repeated for a  rigid circular string  with two equal 
spins  in $AdS^5$ and orbital momentum  $J=J_1$ in $S^5$. The solution is \ci{ft2,art} 
 \be
&&Y_0 + iY_5 = \sqrt{1+2r^2}  \  e^{i \k \tau} \ , \   \  
Y_1 + iY_2 = r \  e^{i ( w \tau +  \s) } \ , \  \
Y_3 + iY_4 = r \  e^{i ( w \tau -  \s) } \ , \ \ \  \\ 
&&  
X_1+i X_2=  e^{i \nu \tau}   \  ,\ \ \ \ \ \  w^2=\kappa^2+1  \ , \ \ \ \ \ \ \ 
\kappa^2 = 4 r^2+\nu^2
\ ,  \la{j}\\
&& \E_0=(1+2r^2)\kappa = \kappa+\frac{2\kappa\S}{\sqrt{1+\kappa^2}}~, ~~~~~~~~~~
\S= \S_1=\S_2=r^2w~, \ \ ~~~~\J=\nu  \ . 
\ee
The parameter $\kappa$ is determined from the conformal gauge condition 
  \be
  \kappa^2=\frac{4}{\sqrt{1+\kappa^2}}\S+\J^2 \ ,
  \ee
and 
leads to the following ``short'' string expansion of the 
 classical energy  ($E_0 = \sql \E_0$):
\be
\E_0 = 2\sqrt{\S}\ \Big( 1 +\S +\frac{\J^2}{8\S }+ ... \Big) \ . \la{eca}
\ee
The leading 1-loop correction to the energy  for  $J=0$ was 
$- \sqrt{\S}$ \ci{rt,ptt2}  and, as in the  previous  section, 
 it should be the same  for small $\J$ as
  well.\foot{The present 
 solution $(E, S_1,S_2; J)$  is related to the $(E; J_1,J_2,J_3)$
 circular solution by an analytic continuation so the fluctuation spectra should be similar.} 
%
The constant shift $b_0$  was zero   for $J=0$ \ci{rt}  but now we expect to 
find $b_0=2$ for $J=2$.
  We thus  end up with  
\be\la{isa}
E=E_0 + E_1 =  2\sqrt{\sqrt{\lambda}  S }\ 
\Big[1+\frac{1}{\sqrt{\lambda}} \Big( S +  
 { J^2\ov 8 S}  - \ha \Big) +  \OO( { 1\ov (\sql)^2  }) 
\Big]   + b_0
\ . \ee
  As for the small circular string in $S^5$, in the flat space limit the  state 
   on the first excited level  associated to the  string
  with two equal spins in $AdS_5$ appears to have only two excited
   oscillators, i.e.  the  corresponding state 
   should  have
    $S=S_1=S_2=1$. Then, for $J=2$,   the dual state  
  should  be in representation $[0,2,0]_{(1,0)} $.
   As in the previous example, there  is   just  one such  state 
   in the Konishi multiplet  table,  at level $ \D_0=6$.
   The resulting  value of $b_1$  is the same as in \rf{v}
   \be 
   b_1=
   2 \Big( S + { J^2 \ov 8  S } - \frac{1}{ 2}\Big)_{{S=1,J=2}} =2 \ . 
   \la{vv}
   \ee

\subsection{ Small circular  spinning  string with $S=J_1$  and  $J_2 \not=0$  
\label{SeqJ1}}

  Next, let us consider a $J=J_2 \not=0$   generalization of  the rigid circular solution 
 with one spin in $AdS_5$ and one angular momentum in $S^5$ discussed  in \ci{rt}
   \be \la{mm}
&&Y_0 + iY_5 = \sqrt{1+r^2}  \  e^{i \k \tau} \ , \ \ \  \  \ \ \,
Y_1 + iY_2 = r \  e^{i ( w \tau +  \s) } \ , \ \ \  \ \  \  \ \   w^2 = \k^2 +1 \ , \\ 
&& X_1 + i X_2  = a \  e^{i ( w'\tau -  \s) } \   , \ \ \ \ \ \ 
X_3+i X_4 = \sqrt{1 -  a^2} e^{i \nu \tau}   \  , \ \ \ \  \ \!w'^2 = \nu^2 +1 \  , \\ 
&&
\kappa^2 -\nu^2 = 2r^2+2a^2
~~~~~~~~~~~~~~~~~~
r^2w=a^2w'\ , \\
&&
\E_0 =\kappa+\frac{\kappa \S}{\sqrt{1+\kappa^2}}
~,~~~\S=r^2w=a^2w'=\J_1~,~~~~ \ \ \  \J_2=(1-a^2)\nu \ . \la{p}
\ee
The parameters $\kappa$  and  $\nu$  may be found by solving the equations
\be
\kappa^2-\nu^2 = \frac{2\S}{\sqrt{1+\kappa^2}} +\frac{2\S}{\sqrt{1+\nu^2}}  
~, \ \ \ \ ~~~~~~~
\J_2=\nu-\frac{\nu \S}{\sqrt{1+\nu^2}} \ . 
\ee
Clearly, $\kappa^2$ is  a series in $\nu^2 \ll 1 $; since,  moreover,  
$\kappa\ne 0$ at $\nu=0$, it follows that $\kappa$ is also a series in 
$\nu^2=\J^2_2+...$. One  finds  that the classical energy is
\be
\E_0 = 2\sqrt{\S}\ \Big(1 +\ha \S +\frac{\J^2_2}{8\S}+... \Big) \ . 
\ee
The  leading 1-loop correction to the energy for $J_2=0$ was vanishing 
in \ci{rt},  and, as we shall argue in sec.\ref{argument} the same  should be true also 
for non-zero $J_2 \ll \sql$.  Therefore, we  find that  
\be\la{isaa}
E=E_0 + E_1 =  2\sqrt{\sqrt{\lambda}  S }\ 
\Big[1+\frac{1}{\sqrt{\lambda}} \Big( \ha S +   { J^2_2\ov 8 S}  \Big) +  \OO( { 1\ov (\sql)^2  }) 
\Big]   + b_0
\ . \ee
Similarly to the other two circular string solutions, to get a state on the 
first excited level we should   choose
$S=J_1=1$.  
 Then for   
$J_2=2$  the corresponding  representation is (relabeling $J_1$ and $J_2$)
 $[1,1,1]_{(\ha,\ha)}$. 
 Unlike the two previous 
 examples,   there  exist several states in the Konishi multiplet  table which transform 
in this representation (at levels $\Delta_0=6$ and $\Delta_0=4,8$). Since in the flat space
limit this state happens to combine
 with the previous two circular string states into a single 
Lorentz-invariant multiplet, we expect that it 
corresponds to a level $\Delta_0=6$  state.
   The resulting $b_1$  takes again the same value:
   \be 
   b_1=
   2 \Big(  \ha  S + { J^2_2 \ov 8  S } \Big)_{{S=1,J_2=2}} =2   \ . 
   \la{vvv} 
   \ee

  \subsection{
  On the independence of  the leading 1-loop correction to string energy 
   on $J \ll \sql$ \label{argument} }
  
  One  may   give a general argument suggesting that 
  if a  semiclassical angular 
 momentum  $J=\sql \J \ll \sql  $ is added to a classical solution
  carrying other charges (of the same order 
 or larger) in the quantum corrections to the string 
  energy its presence will first show up 
 at order $\J^2 = { J^2 \ov (\sql)^2}$. 
 The leading   $  1 \ov \sql$ order  (for fixed $J$)  correction 
  is then not affected 
 by $J$ as  it happened in the examples discussed above. 
 
 The $\J \rightarrow  -\J$ symmetry of the string  Lagrangian expanded around a 
 solution
 of the type described above implies that  it depends on $\J$ only through 
  $\J^2$. This need not,  however,    immediately  apply to  quantum corrections
 to the energy or other charges. For example, for special fluctuation 
 modes  the corresponding solutions of the 
 characteristic equations may  depend,  in fact,  on $|\J|$  suggesting that 
 such terms   may potentially appear  in the 1-loop   correction to the energy.
 
 One may  make  an  assumption    that  if   $\J$ is added 
 to a non-trivial  classical solution, then observables should be 
  continuous functions of 
 $\J$, i.e. that the limit $\J\rightarrow 0$ should give the same answer whether 
 one approaches the  $\J=0$ point from $\J>0$ or from $\J<0$.\foot{We do not assume, 
 however, that this $\J\to 0$ limit reproduces the
 value of the observable computed directly at 
 $\J=0$. The energies of folded spinning string solutions discussed in the next section  
 are examples of cases when these two quantities are different.}
 

Suppose we   consider  a string partition function as  function 
of  parameters  of the classical solution  and  evaluate the derivative of 
it  with respect to $\J$. Potential terms containing $|\J|$ would make this 
derivative discontinuous at $\J=0$ as ${d|\J|\ov d\J }= {\rm sgn}(\J)$.
As discussed in  \ci{rt0712}, the semiclassical parameter $\J$ may be given the 
interpretation of the chemical potential for the angular momentum $J$. The 
conformal gauge constraint relates this chemical potential to 
chemical potentials
for other charges and for the energy, i.e. 
$
\kappa=\kappa(\J^2,...).
$
Then according to  \ci{rt0712} 
$
\frac{d}{d\J}\ln Z = \frac{d\kappa}{d\J}\langle E\rangle - \langle J\rangle~.
$
Since $\kappa$ depends on $\J^2$ and $\langle E\rangle$ and  
$\langle J\rangle$ are assumed to be 
continuous functions at $\J=0$, it then follows that 
$\frac{d}{d\J}\ln Z$ must also be continuous, in particular,   $\ln Z$ cannot contain 
terms proportional to $|\J|$. 
Higher odd powers of $|\J|$ are not forbidden by this 
argument, but they would  not contribute  at leading  $ 1 \ov \sql$  order 
($|\J|^k= { |J|^k \ov  (\sql)^k}$). 

The assumption of  continuity of  observables as $\J\rightarrow 0$ is potentially subtle, 
and it is possible  that  the limit $\J\to 0$ may or may not be analytic, depending on a 
solution in question. Since, however, 
 $\J$ is turned on in the presence of other  classical charges, 
 one may expect  this non-analyticity to be weaker than,  e.g., in  the example 
discussed in  Appendix A in \ci{rt}. There   we found different results  when  
taking  $\J\to 0$ before or after  doing summation over infinite number of quantum 
modes. Unlike the setup discussed here, in that case the
classical solution was becoming trivial in the  limit $\J\to 0$.

\renewcommand{\theequation}{3.\arabic{equation}}
 \setcounter{equation}{0} 
 
 \section{Examples of semiclassical string states
  dual to\\  $\D_0=4$ states in Konishi multiplet } 
 
 Similar observations  
 regarding the generalization to nonvanishing orbital  momentum $J$ in $S^5$ 
 go through in the case of more complicated folded 
 spinning string solutions. 
 
\subsection{Folded string with spin $S$   and orbital momentum $J$ }

Let us start with   the case of folded spinning string with $J=0$. 
The ``short''-string (i.e. $\S = { S \ov \sql } \ll 1$)  expansion of the 1-loop corrected 
energy  of this solution was  considered  in \ci{tir} (using near-flat space expansion)
and in \ci{bef} (by directly diagonalizing the  corresponding fluctuation problem for any $\S$).
The resulting expression for the energy  was found to be \ci{tir,bef}
\be
 && E=E_0 + E_1=  \sqrt{2\sqrt{\lambda} S}\ 
\Big[1+\frac{1}{\sqrt{\lambda}} \Big( {{ 3 \ov 8}}   S 
 + a^{(0)}_{01}\Big) 
+... \Big]  + b^{(0)}_0  \ , \la{ai}  \\
&&  a^{(0)}_{01} = 3 - 4 \ln 2 \ , \ \ \ \ \ \ \  \ \ \ \  b^{(0)}_0 =1   \ . \la{io}
\ee
The  generalization of  \rf{ai}  to the case of the folded spinning string 
with  non-zero  angular momentum $J\ll \sql $ in $S^5$ \ci{ft1} is given by \ci{bet,rt}
\be
  E=E_0 + E_1=  \sqrt{2\sqrt{\lambda} S}\ 
\Big[1+\frac{1}{\sqrt{\lambda}} \Big( {{ 3 \ov 8}}   S  +   {J^2 \ov 4S} +  a_{01}\Big) 
+... \Big]  + b_0  \ . \la{aii}  
\ee
A direct calculation of the 1-loop  coefficients $a_{01}, b_0$ 
starting with the string fluctuation Lagrangian appears  to be 
non-trivial due to mixing of the fluctuation modes \ci{ft1,bet}.
%

The computation of  $a_{01}, b_0$ in \rf{aii}  was first performed in   \ci{gr}. 
It  started with the folded spinning string solution with 
$J={ \J \ov \sql }  \not=0$  \ci{ft1,bfst}
and  used  the algebraic curve approach \ci{curve} to extract the fluctuation spectrum. 
The resulting 1-loop correction to the energy 
was  then expanded first in $\J \ll 1 $ and then in $\S \ll 1 $  with the result being 
 \rf{ai}
with  coefficients different from \rf{io} \foot{The results of \ci{gr} where 
numeric, implying that $a_{01}  = -0.25...$. In \ci{rt}  it was 
assumed that $a_{01} $ is actually equal to $ -{ 1 \ov 4}$, as that led to a
consistency between   values of 1-loop  corrections for other   string states 
 that may appear on the first excited level.
 Very recently  the  expressions in \rf{o}   were derived analytically \ci{gron}
 using the same  algebraic curve method as in \ci{gr}. 
 We  thank  N. Gromov for informing us  about this  result  prior to its publication.}
\be  a_{01}   = -{ 1 \ov 4}  
\ , \ \ \ \ \ \ \  \ \ \ \  b_0  =0   
\ . 
\la{o}
\ee
The  difference between  \rf{io}  and \rf{o}   implies that in this case 
the limit $\J \to 0$ is non-analytic.\foot{
Possible reasons for this non-analyticity are the non-trivial mixing of fluctuation 
modes  which was absent in the $\J=0$ case and the difference in the number
of massless fluctuation modes between the $\J=0$ and $\J\not=0$ cases.
}  

Starting with \rf{aii},\rf{o}  we then get a state on the first excited string level 
if we set $S=2$. Assuming also $J=J_1=2$   we can then associate to 
this   state  a state $[0,2,0]_{(1,1)} $ in the Konishi multiplet     with $ \D_0=4 $
which is  the familiar Konishi descendant in the  $sl(2)$ sector.
The value of  $b_0  =0 $ is consistent  with $ \D_0=4 $.
Then   the value of  $b_1$  in \rf{fow}   is found to be 
the same as in the examples  in the previous section and  in agreement   with \ci{gkv,fro}:
 \be 
 b_1= 2 \Big( {{ 3 \ov 8}}   S  +   {J^2 \ov 4S} - { 1 \ov 4}   \Big)_{{S=J=2}} =2  \ . 
  \la{f} \ee

\subsection{Folded string with spin $J_1$  and orbital momentum $J_2$  }

One may  repeat  the above  discussion for the folded 2-spin string solution 
in $S^5$ \ci{ftp}.  For the  string  with  c.o.m.  at rest ($J_2=0$)   
one finds \ci{bec} (cf. \rf{ai},\rf{io})
\be
 && E=E_0 + E_1=  \sqrt{2\sqrt{\lambda} J_1}\ 
\Big[1+\frac{1}{\sqrt{\lambda}} \Big( {{ 1 \ov 8}}   J_1
 + a^{(0)}_{01}\Big) 
+... \Big]  + b^{(0)}_0  \ , \la{aai}  \\
&&  a^{(0)}_{01} = {2} - 4 \ln 2 \ , \ \ \ \ \ \ \  \ \ \ \  b^{(0)}_0 =2   \ . \la{iao}
\ee
Switching on non-zero $J_2$  we get (cf. \rf{aii})
\be
  E=E_0 + E_1=  \sqrt{2\sqrt{\lambda} J_1}\ 
\Big[1+\frac{1}{\sqrt{\lambda}} \Big( {{ 1 \ov 8}}   J_1  +   {J^2_2 \ov 4J_1} +  a_{01}\Big) 
+... \Big]  + b_0  \ . \la{aaii}  
\ee
Since the  two folded string solutions ($(E,S;J)$ and $(E; J_1,J_2)$)
are closely related when all spins are non-zero (cf. \ci{bfst})
 the corresponding  
results for the 1-loop corrections in the ``short''-string (near-flat  space) 
limit  should  be similar.  We may thus  conjecture, following \ci{rt}, that $a_{01}$ here 
should be the 
same as \rf{o} up to the opposite sign  (reflecting the opposite sign 
of the curvature of $S^3$ as compared to $AdS_3$)\foot{Note a similar
 opposite sign of 1-loop corrections in \rf{py} and in \rf{isa}.}
\be  a_{01}   = { 1 \ov 4}  
\ , \ \ \ \ \ \ \  \ \ \ \  b_0  =0   
\ . \la{ou}
\ee
Here  the limit $\J_2 \to 0$ in the 1-loop correction to energy 
should  again  be  non-analytic.

Considering the  case with $J_1=J_2=2$ we then  get a state on the first excited string level 
with the corresponding representation 
  $[2,0,2]_{(0,0)} $  present in the Konishi multiplet  at  level  $ \D_0=4 $.
  The dual operator should be  the $su(2)$ sector descendant of the Konishi operator.
  The corresponding value of $b_1$  in \rf{fow} that follows from \rf{aaii}  
  is once again  equal to 2 
 \be 
 b_1= 2 \Big( {{ 1 \ov 8}}   J_1  +   {J^2_2 \ov 4J_1} + { 1 \ov 4}   \Big)_{{J_1=J_2=2}} =2  \ .
 \la{ff}  \ee

\subsection{Comments on other states}

 As in \ci{rt} one may also consider a string folded and spinning (with spins $S$ and $J_1$) 
  in both $AdS_5$ and $S^5$.
   It was conjectured in \ci{rt} that in this case the leading 
   1-loop contributions to $a_{01}$  cancel out (see \rf{o}  and \rf{ou}; cf. also \rf{isaa})
    so that the leading
   correction to the energy is given just by  the classical term. 
   Adding  extra  orbital momentum  in $S^5$, i.e.
   generalizing the $S^5$ part to 
   $(J_1,J_2)$  folded spinning string  we then get as in \ci{rt} 
     \be\la{bkii}
E=E_0+E_1= \sqrt{2\sqrt{\lambda} (S+J_1) }\ 
\Big[1+\frac{1}{\sql} \Big( { 3 \ov 8} S   +  { 1 \ov 8} J_1 +  
 {J_2^2\ov 4(S + J_1) } 
   \Big) + ...  \Big]+b_0
   \ee
  for $S+J_1=2$  we then get  a state  on the first excited  level;  with 
  $J_2=2$  the corresponding representation  is 
  (interchanging $J_1$ and $J_2$)   $[1,1,1]_{(\ha,\ha)}$, i.e. the same as in the
   case of the circular $(S,J_1; J_2)$ solution in sec.~\ref{SeqJ1}. 
   There is no contradiction as  there are several 
   such  $[1,1,1]_{(\ha,\ha)}$  states in the Konishi multiplet table 
   (e.g. four at level $\D_0=6$).  
  Once again, 
  \be 
 b_1= 2 \Big[ {{ 3 \ov 8}} S +   {{ 1 \ov 8}} J_1 
 +   {J^2_2 \ov 4(S+J_1)}    \Big]_{{S=J_1=1, J_2=2}} =2  \ .
 \la{fff}  \ee 
  The transcendental  values of  the 1-loop coefficients in \rf{io},\rf{iao}
 for the single-spin folded string solutions imply  that  corresponding states should 
 not belong to the Konishi multiplet. 
 %
 %
 One may wonder if such  semiclassical states cannot actually be interpolated to 
 true quantum string states.

 Another open question is the identification of the string states associated  to 
 singlet (spinless) states in the Konishi multiplet.
 A natural guess could be that  they  are  small 
  pulsating strings in $AdS_3$ or $R \times S^2$ \ci{min,bec}.
  However, the corresponding  1-loop coefficient $a_{01}$ 
   contains  again $\ln 2$ terms \ci{bec}.    By analogy with the folded string case  
   one may conjecture 
   that first  adding  a non-zero  orbital momentum $J$  and then expanding in $J \ll \sql$ 
    will make $a_{01}$ rational. 
   For example, for a string  pulsating in $AdS_5$ with 
   the oscillation number $N\ll \sql $ and 
    orbital  momentum $J\ll \sql $ in $S^5$  one finds as in  \ci{bec}  and above (cf. \rf{rty})
   \be
  E=E_0 + E_1=  \sqrt{2\sqrt{\lambda} N }\ 
\Big[1+\frac{1}{\sqrt{\lambda}} \Big( {{ 5 \ov 8}}  N  +   {J^2 \ov 4N} +  a_{01}\Big) 
+... \Big]  + b_0  \ . \la{aaiiy}  
\ee
The choice of $N=2$ and $J=2$  then gives a  state on the  first excited string level.
 A candidate dual  state  in the Konishi multiplet 
  is  $ [0,2,0]_{(0,0)} $ at level $\D_0=4$.  
  To reproduce 
   the same universal   value $b_1=2$ as above we would  then need  $a_{01}=  - { 3 \ov 4}$
   (instead of  $a^{(0)}_{01}= {5 \ov 2} - 4 \ln 2$ found in \ci{bec} by 
   starting directly with the solution having 
   $J=0$). 
 
  The value of the  string center-of-mass 
 orbital   $S^5$ momentum $J$ does not influence 
 the  value of the string excitation level which is determined by  spins  and 
 oscillation numbers  related to extended nature of the string (cf.~\rf{rty}).
 The states with $N=2$ in \rf{rty} but $J > 2$ will still be on the first excited level.
 In general, they will not belong to the Konishi multiplet 
 but rather to its Kaluza-Klein descendants (see \ci{bia}).

 A non-vanishing value of the orbital angular momentum $J$ is crucial
 for relating some of the classical solutions we discussed to members of the Konishi 
 multiplet. For example, for $J=0$, the circular string solution with one spin in 
 $AdS_5$ and one in $S^5$ discussed in section~\ref{SeqJ1} 
 should have 
 $S=1=J_1$  (to represent a state on the first excited string level). It should also correspond   
 to a gauge-theory operator  in the  $sl(2)$ sector, 
 but  the minimal R-charge in the $sl(2)$  sector is $J_1=2$, so 
 such $S=1=J_1$ state should be excluded. 

 Solutions with vanishing orbital angular momentum $J$  have a natural 
 place on higher string levels, with level determined by values of 
 the $AdS_5$ and $S^5$ spins.
 %
 Choosing the spins so that the  solutions correspond to the second excited string
  level ($N=4$ in \rf{rty}) 
  we find from the eqs. \rf{py}, \rf{isa}, \rf{isaa} the  following values of $b_1$ in the 
  analog of \rf{fow} ($E= \sqrt{ 2 N} \fl +  b_0 + { b_1 \ov \fl} + \dots $): \ \ 
$
b_1(J_1=J_2=2)=\sqrt{2};\ \ 
b_1(S_1=S_2=2)=3\sqrt{2};\ \ 
b_1(S=J_1=2)=2\sqrt{{2}}$. 
From   the  eqs. \rf{aii}, \rf{aaii}, \rf{bkii} we get:\foot{
The Konishi multiplet also contains a state in the representation 
$[0,0,0]_{(2,2)}$, which has $S=4$. It should correspond to a string 
state on the first excited level with 2 units of spin  carried by the  two string 
oscillators and the other two  units of spin carried by the graviton vacuum 
state.} \ \  
 $
b_1(S=4)=\frac{5}{2}{\sqrt 2};\ \ 
b_1(J=4)=\frac{3}{2}{\sqrt2};\ \ 
b_1(S=J_1=2)=2\sqrt{{2}}.
$
These different values of the coefficient $b_1$ for states on the  same  string 
level are  not in  contradiction with supersymmetry.
 Indeed, the corresponding level in the  flat space string spectrum contains 81 
physical long multiplets, each of them generating a KK tower in \adss \cite{bia}. 
While the leading order terms in the energy of all such states are equal
 to $2\sqrt{2} \fl $, the order ${1 \ov \fl}$ corrections can be different.

 Finally, it goes without saying that 
  the construction of a systematic quantization of \adss 
  superstring in the perturbative large tension expansion remains an 
  important open problem. Some related recent work  this direction appeared in \ci{ja,plef}. 
   The consistent 
   results found  in the semiclassical approach should provide an important guidance.

\section*{Acknowledgments }

We thank  M. Beccaria, N. Gromov  and   A. Tirziu 
   for  many  useful  discussions.
   The  work of RR  was supported by the US Department of Energy under contract
DE-FG02-201390ER40577 (OJI), 
the US National Science Foundation under grant PHY-0855356 and 
the A. P. Sloan Foundation.
     
We would like to  thank N. Gromov   for informing us  about  
 a   forthcoming  paper \ci{gron} 
in which eq.\rf{o} is derived  analytically 
 and the conclusion   that the semiclassical approach  applied to the case of the
folded spinning string with $S=J=2$  leads to
 the same value $b_1=2$   as in \ci{gkv}  is   reached independently.
 We also thank B. Vallilo for  sending us a draft of \ci{val}
 which  also  claims, using a different method,
   that $b_1=2$    for a different 
  state  (with zero $S^5$ momentum) on the first excited string level.
 
%
%



\




\appendix
\section*{Appendix: 1-loop correction to energy of 
3-spin solution of sec. \ref{J1J2J3} }

\refstepcounter{section}
\def\theequation{A.\arabic{equation}}
\setcounter{equation}{0}


Here we present some  details of the calculation of the 1-loop 
correction to the energy of the 3-spin solution \cite{ft2,ft3} in the small-string 
regime discussed in section \ref{J1J2J3}.
In this  case  $\J_1=\J_2=\J' \ll  1  $.
 The equations of motion and the conformal 
gauge constraints may be easily solved and the resulting  expansions in 
 small $\J$  for fixed $\J'$ are 
 \be
\label{nu}
\nu&=&\frac{\J}{1 - 2 \J'}\Big[     1        -  \frac{\J' }{(1- 2 \J')^3}   \J^2     +
 \frac{3\J'(1+2\J')}{4(1-2\J')^6}\J^4+{\cal O}(\J^6)   \Big] \ , 
\\
{\cal E}_0=\kappa&=&
2\sqrt{\J'}\Big[1+\frac{\J^2}{8(1-2\J')\J'}-\frac{(12\J'{}^2-4\J'+1)\J^4}{128(1-2\J')^4\J'{}^2}
+{\cal O}(\J^6)\Big]  \ , 
\label{EJltJp} 
 \ee
leading to the  expansion of the classical energy in 
\rf{ec}.\foot{Let us  mention
that starting with $1 \ov (\sql)^{2}$ order, the
energy  depends not just on $J^2\ov J'$, but independently on $J$ and $J'$
(e.g.  one finds terms like $J^2\ov 4\lambda$).}

The 1-loop calculation may  proceed either by evaluating the frequencies of the 
 physical fluctuations around the solution and constructing 
$
E_1=\frac{1}{2\kappa}\sum_{n=-\infty}^{+\infty}(-)^{F_n}\omega_n~,
$
or by directly evaluating the determinant of the quadratic fluctuation operator. 
We will  use  the fluctuation frequencies found in \cite{ft3} and expand 
them in small  $\J$.
 Then  $E_1$  has the following expansion 
\be\la{er}
E_1 =\frac{1}{2\kappa}\sum_{n=-\infty}^{+\infty}(-)^{F_n}\omega_n=\frac{1}{\kappa}
\left[f_0(\J')
+f_1(\J')\J+f_2(\J')\J^2+{\cal O}(\J^3)\right]~~.
\ee
The analyticity properties of the functions $f_i(\J')$ determine the
order in the $1\ov \sql$ expansion (for fixed $J,J'$) to which they contribute. In general,
these functions appear to be analytic, or at most have  simple poles in
$\sqrt{\J'}$.  In particular, the third term, proportional to $\J^2$, is of 
too high an order in the large $\sql$ expansion to be relevant for  the leading 
${1\ov\sql}$ correction that we are concerned with here.
The argument outlined in section  2.4 suggests 
that $f_1(\J')=0$. We will  see that this is indeed the case. 

As  found in \cite{ft3}, apart from two massless modes and four bosonic modes with 
$
\omega_n=\sqrt{n^2+\kappa^2}
$
the characteristic frequencies are the roots of the ``bosonic'' and ``fermionic'' 
polynomials 
($\Omega\equiv \omega^2$  and $q \equiv  {2  \J'}/{\sqrt{1 + \nu^2}} $)
\begin{eqnarray}
\lefteqn{\nonumber B_8(\Omega)=\Omega^4+\Omega^3 \left(-8-4
n^2+20 q-8 \kappa ^2\right)}\\
&&\nonumber+\Omega^2(16+8 n^2+6 n^4-80 q
-36n^2 q+96 q^2+32 \kappa ^2+16 n^2 \kappa ^2
-80 q \kappa ^2+16 \kappa^4)\\
&&\nonumber+\Omega (-32 n^2+8 n^4
-4 n^6+96 n^2 q+12 n^4 q-96 n^2 q^2
-32 n^2 \kappa ^2-8 n^4 \kappa ^2+48 n^2 q \kappa ^2)\\
&&+16 n^4-8 n^6+n^8-16 n^4 q+4 n^6 q \ , 
\nonumber \\
\lefteqn{\nonumber F_8(\Omega)=2\Omega^4+\Omega^3(-8-12\kappa^2
-8n^2+20q)}
\\
&&\nonumber+\Omega^2(12+28\kappa^2+18\kappa^4+8n^2
+28\kappa^2n^2+12n^4 
-52q-64\kappa^2q-36n^2q+59q^2)\\
&&\nonumber+\Omega(-8-20\kappa^2-20\kappa^4-8\kappa^6
+8n^2+8n^2\kappa^2-20\kappa^4n^2+8n^4-20\kappa^2n^4\\
&&\nonumber~~~~~
-8n^6+44q+80\kappa^2q+44\kappa^4q-24n^2q
+32\kappa^2n^2q+12n^4q-78q^2\\
&&\nonumber~~~~~
-79\kappa^2q^2+2n^2q^2+45q^3)
\\&&\nonumber 
+2+4\kappa^2+2\kappa^4-8n^2-4\kappa^2n^2-4\kappa^4n^2
 +12n^4-4\kappa^2n^4+2\kappa^4n^4\\
&&\nonumber
-8n^6+4\kappa^2n^6+2n^8-12q-16\kappa^2q-4\kappa^4q
+28qn^2+16\kappa^2n^2q\\
&&\nonumber
+4\kappa^4n^2q-20n^4q+4n^6q+27q^2+21\kappa^2q^2
+2\kappa^4q^2-30n^2q^2\\
&&-11\kappa^2n^2q^2+11n^4q^2-27q^3-9\kappa^2q^3
+9n^2q^3+{\te {81\ov 8}}q^4~.
\end{eqnarray}
A closed-form expression for the characteristic frequencies does not seem to exist
but one may readily find  their expansion in small $\J$. 
 It turns 
out that
 it is more convenient to use $\nu\propto \J$ (cf. eq.(\ref{nu}))
as an expansion parameter.  We will also limit ourselves to terms 
of order $ \nu$,
as higher order terms scale at least as  $1 \ov (\sql)^2$  for fixed $J$. 

The expansion of the squares of the four nontrivial bosonic frequencies is
\be
\Omega^b_1&=&n^2-\frac{2\sqrt{\J'}n^2}{(n^2+2\J'-n^2\J'-1)^{1/2}}\nu
+{\cal O}(\nu^2)\ , 
\cr
\Omega^b_2&=&n^2+\frac{2\sqrt{\J'}n^2}{(n^2+2\J'-n^2\J'-1)^{1/2}}\nu
+{\cal O}(\nu^2)\ , 
\nonumber\\
\Omega^b_3&=&4+n^2-4 \J'-4\sqrt{\J'{}^2-n^2 \J'{}+n^2}
+0\times\nu+{\cal O}(\nu^2)\ , 
\nonumber\\[1pt]
\Omega^b_4&=&4+n^2-4 \J'+4\sqrt{\J'{}^2-n^2 \J'{}+n^2}
+0\times \nu+{\cal O}(\nu^2)\nonumber
\ee
and the expansion of the squares of the four nontrivial fermionic frequencies is
\be
\Omega^f_1  \!\!&=&\!\!
1+\J'+n^2-2\sqrt{n^2(1-\J')+\J'}
+\frac{\sqrt{\J'} 
\left(n^2-\sqrt{n^2(1-\J')+\J'}+1\right) }
{\sqrt{n^2(1-\J')+\J'}}\nu+{\cal O}(\nu^2)\cr
\cr
\Omega^f_2  \!\!&=&\!\!
1+\J'+n^2+2\sqrt{n^2(1-\J')+\J'}
-\frac{\sqrt{\J'} 
\left(n^2+\sqrt{n^2(1-\J')+\J'}+1\right) }
{\sqrt{n^2(1-\J')+\J'}}\nu+{\cal O}(\nu^2)\cr
\cr
\Omega^f_3  \!\!&=&\!\!
1+\J'+n^2-2\sqrt{n^2(1-\J')+\J'}
-\frac{\sqrt{\J'} 
\left(n^2-\sqrt{n^2(1-\J')+\J'}+1\right) }
{\sqrt{n^2(1-\J')+\J'}}\nu+{\cal O}(\nu^2)\cr
\cr
\Omega^f_4  \!\!&=&\!\!
1+\J'+n^2+2\sqrt{n^2(1-\J')+\J'}
+\frac{\sqrt{\J'} 
\left(n^2+\sqrt{n^2(1-\J')+\J'}+1\right) }
{\sqrt{n^2(1-\J')+\J'}}\nu+{\cal O}(\nu^2)\nonumber 
\ee
As discussed in \cite{rt}, when carrying out the frequency sum it is important to
isolate the modes that have a non-analytic dependence in $\J'$. This is the case
for the $n=0, \pm 1$ modes, for which some of the $\Omega=\omega^2$  or some of 
the frequencies $\omega$ acquire $\sqrt{\J'}$-dependent terms. We will also isolate 
the $n=\pm 2$ mode. Carrying out the sum and expanding at small $\J'$ leads to
the following results   for the functions in \rf{er}
\be
\label{f0}
f_0(\J')&=&\Big(\frac{7}{3}-\frac{1}{3}\Big)\J'
+\Big(-\frac{3445}{432}+\frac{1}{432}
[3121-2592\zeta(3) ]\Big)\J'{}^2+{\cal O}(\J'{}^3) 
\cr
&=&2\J'-\frac{3}{4}\Big[ 1+8\zeta(3) \Big]\J'{}^2+{\cal O}(\J'{}^3) 
\\
f_1(\J')&=& 0   \ , \ \ \ \ \ \ \ \   \ \ \ \ \ \ f_2(\J')={\cal O}(\J')~~.
\ee
In each parenthesis on the first line of (\ref{f0}) the first number is the
contribution of the $n=0,\pm1,\pm2$ modes and the second number is the
contribution of the other modes.
 The vanishing of the function $f_1(\J')$ 
implies  the absence of terms linear in $\J$, in agreement with the general argument 
in sec.~\ref{argument}.

Extracting  the leading 
correction to the energy of 
the small 3-spin circular 
string in the limit $\J \ll 1$ 
we find   that it is independent of  $J$
\be
E_1={\sqrt{\sql J'}}\  \Big[  { 1 \ov \sql }  + {\cal O}({ 1 \ov (\sql)^2}) \Big] \ .
\ee
Combining this with the tree-level energy in   (\ref{EJltJp})
we find that the the total energy is given by \rf{py}. 

\begin{table}[htb] 
\centering
{\footnotesize
\begin{tabular}{|c|l|} \hline
     $\D_0$ & $[p_1, q, p_2]_{(s_L, s_R)}=
[J_2-J_3, J_1-J_2, J_2+J_3]_{( {S_1+S_2 \ov 2}, {S_1-S_2 \ov 2})}$\\ \hline\hline
     $2$& $
     [0,0,0]_{(0,0)}
     $ \\ \hline $
     \dDelta+\frac{1}{2}$& $
     [0,0,1]_{(0,\frac{1}{2})}
     +[1,0,0]_{(\frac{1}{2},0)}
     $ \\ \hline $
     \dDelta+1$& $
     [0,0,0]_{(\frac{1}{2},\frac{1}{2})}
     +[0,0,2]_{(0,0)}
     +[0,1,0]_{(0,1)+(1,0)}
     +[1,0,1]_{(\frac{1}{2},\frac{1}{2})}
     +[2,0,0]_{(0,0)}
     $ \\ \hline $
     \dDelta+\frac{3}{2}$& $
     [0,0,1]_{(\frac{1}{2},0)+(\frac{1}{2},1)+(\frac{3}{2},0)}
     +[0,1,1]_{(0,\frac{1}{2})+(1,\frac{1}{2})}
     +[1,0,0]_{(0,\frac{1}{2})+(0,\frac{3}{2})+(1,\frac{1}{2})}
     +[1,0,2]_{(\frac{1}{2},0)}
     $\\& $
     +[1,1,0]_{(\frac{1}{2},0)+(\frac{1}{2},1)}
     +[2,0,1]_{(0,\frac{1}{2})}
     $ \\ \hline $
     \dDelta+2$& $
     [0,0,0]_{(0,0)+(0,2)+(1,1)+(2,0)}
     +[0,0,2]_{(\frac{1}{2},\frac{1}{2})+(\frac{3}{2},\frac{1}{2})}
     +[0,1,0]_{2(\frac{1}{2},\frac{1}{2})+(\frac{1}{2},\frac{3}{2})+(\frac{3}{2},\frac{1}{2})}
     +[2,0,2]_{(0,0)}
     +[2,1,0]_{(0,1)}
     $\\& $
     +[0,1,2]_{(1,0)}
     +[0,2,0]_{2(0,0)+(1,1)}
     +[1,0,1]_{(0,0)+2(0,1)+2(1,0)+(1,1)}
     +[1,1,1]_{2(\frac{1}{2},\frac{1}{2})}
     +[2,0,0]_{(\frac{1}{2},\frac{1}{2})+(\frac{1}{2},\frac{3}{2})}
     $ \\ \hline $
     \dDelta+\frac{5}{2}$& $
     [0,0,1]_{(0,\frac{1}{2})+(0,\frac{3}{2})+2(1,\frac{1}{2})+(1,\frac{3}{2})+(2,\frac{1}{2})}
     +[0,0,3]_{(\frac{3}{2},0)}
     +[0,1,1]_{3(\frac{1}{2},0)+2(\frac{1}{2},1)+(\frac{3}{2},0)+(\frac{3}{2},1)}
     +[0,2,1]_{(0,\frac{1}{2})+(1,\frac{1}{2})}
     $\\& $
     +[1,0,0]_{(\frac{1}{2},0)+2(\frac{1}{2},1)+(\frac{1}{2},2)+(\frac{3}{2},0)+(\frac{3}{2},1)}
     +[1,0,2]_{(0,\frac{1}{2})+2(1,\frac{1}{2})}
     +[1,1,0]_{3(0,\frac{1}{2})+(0,\frac{3}{2})+2(1,\frac{1}{2})+(1,\frac{3}{2})}
     $\\& $
     +[1,1,2]_{(\frac{1}{2},0)}
     +[1,2,0]_{(\frac{1}{2},0)+(\frac{1}{2},1)}
     +[2,0,1]_{(\frac{1}{2},0)+2(\frac{1}{2},1)}
     +[2,1,1]_{(0,\frac{1}{2})}
     +[3,0,0]_{(0,\frac{3}{2})}
     $ \\ \hline $
     \dDelta+3$& $
     [0,0,0]_{(\frac{1}{2},\frac{1}{2})+(\frac{1}{2},\frac{3}{2})+(\frac{3}{2},\frac{1}{2})+(\frac{3}{2},\frac{3}{2})}
     +[0,0,2]_{2(0,0)+(1,0)+2(1,1)+(2,0)}
     +[0,1,0]_{3(0,1)+3(1,0)+2(1,1)+(1,2)+(2,1)}
     $\\& $
     +[0,1,2]_{2(\frac{1}{2},\frac{1}{2})+(\frac{3}{2},\frac{1}{2})}
     +[0,2,0]_{3(\frac{1}{2},\frac{1}{2})+(\frac{1}{2},\frac{3}{2})+(\frac{3}{2},\frac{1}{2})}
     +[0,2,2]_{(0,0)}
     +[0,3,0]_{(0,1)+(1,0)}
     $\\& $
     +[1,0,3]_{(1,0)}
     +[1,1,1]_{2(0,0)+2(0,1)+2(1,0)+2(1,1)}
     +[1,2,1]_{(\frac{1}{2},\frac{1}{2})}
     +[2,0,0]_{2(0,0)+(0,1)+(0,2)+2(1,1)}
     $\\& $
     +[2,0,2]_{(\frac{1}{2},\frac{1}{2})}
     +[2,1,0]_{2(\frac{1}{2},\frac{1}{2})+(\frac{1}{2},\frac{3}{2})}
     +[2,2,0]_{(0,0)}
     +[3,0,1]_{(0,1)}
     +[1,0,1]_{4(\frac{1}{2},\frac{1}{2})+2(\frac{1}{2},\frac{3}{2})+2(\frac{3}{2},\frac{1}{2})+(\frac{3}{2},\frac{3}{2})}
     $ \\ \hline $
     \dDelta+\frac{7}{2}$& $
     [0,0,1]_{2(\frac{1}{2},0)+3(\frac{1}{2},1)+(\frac{3}{2},0)+2(\frac{3}{2},1)+(\frac{3}{2},2)}
     +[0,0,3]_{(0,\frac{1}{2})+(1,\frac{1}{2})}
     +[0,1,1]_{3(0,\frac{1}{2})+(0,\frac{3}{2})+4(1,\frac{1}{2})+2(1,\frac{3}{2})+(2,\frac{1}{2})}
     $\\& $
     +[0,1,3]_{(\frac{1}{2},0)}
     +[0,2,1]_{2(\frac{1}{2},0)+2(\frac{1}{2},1)+(\frac{3}{2},0)}
     +[0,3,1]_{(0,\frac{1}{2})}
     +[1,0,0]_{2(0,\frac{1}{2})+(0,\frac{3}{2})+3(1,\frac{1}{2})+2(1,\frac{3}{2})+(2,\frac{3}{2})}
     $\\& $
     +[1,0,2]_{2(\frac{1}{2},0)+2(\frac{1}{2},1)+(\frac{3}{2},0)+(\frac{3}{2},1)}
     +[1,1,0]_{3(\frac{1}{2},0)+4(\frac{1}{2},1)+(\frac{1}{2},2)+(\frac{3}{2},0)+2(\frac{3}{2},1)}
     +[1,1,2]_{(0,\frac{1}{2})+(1,\frac{1}{2})}
     $\\& $
     +[1,2,0]_{2(0,\frac{1}{2})+(0,\frac{3}{2})+2(1,\frac{1}{2})}
     +[1,3,0]_{(\frac{1}{2},0)}
     +[2,0,1]_{2(0,\frac{1}{2})+(0,\frac{3}{2})+2(1,\frac{1}{2})+(1,\frac{3}{2})}
     +[2,1,1]_{(\frac{1}{2},0)+(\frac{1}{2},1)}
     $\\& $
     +[3,0,0]_{(\frac{1}{2},0)+(\frac{1}{2},1)}
     +[3,1,0]_{(0,\frac{1}{2})}
     $ \\ \hline $
     \dDelta+4$& $
     [0,0,0]_{3(0,0)+3(1,1)+(2,2)}
     +[0,0,2]_{3(\frac{1}{2},\frac{1}{2})+(\frac{1}{2},\frac{3}{2})+(\frac{3}{2},
     \frac{1}{2})+(\frac{3}{2},\frac{3}{2})}
     +[0,1,0]_{4(\frac{1}{2},\frac{1}{2})+2(\frac{1}{2},\frac{3}{2})+2(\frac{3}{2},
     \frac{1}{2})+2(\frac{3}{2},\frac{3}{2})}
     $\\& $
     +[0,1,2]_{(0,0)+2(0,1)+2(1,0)+(1,1)}
     +[0,2,0]_{3(0,0)+(0,1)+(0,2)+(1,0)+3(1,1)+(2,0)}
     +[0,2,2]_{(\frac{1}{2},\frac{1}{2})}
     $\\& $
     +[0,3,0]_{2(\frac{1}{2},\frac{1}{2})}
     +[0,4,0]_{(0,0)}
     +[1,0,1]_{(0,0)+3(0,1)+3(1,0)+4(1,1)+(1,2)+(2,1)}
     +[1,0,3]_{(\frac{1}{2},\frac{1}{2})}
     +[0,0,4]_{(0,0)}
     $\\& $
     +[1,1,1]_{4(\frac{1}{2},\frac{1}{2})+2(\frac{1}{2},\frac{3}{2})+2(\frac{3}{2},\frac{1}{2})}
     +[1,2,1]_{(0,0)+(0,1)+(1,0)}
     +[2,0,0]_{3(\frac{1}{2},\frac{1}{2})+(\frac{1}{2},
     \frac{3}{2})+(\frac{3}{2},\frac{1}{2})+(\frac{3}{2},\frac{3}{2})}
     $\\& $
     +[2,0,2]_{(0,0)+(1,1)}
     +[2,1,0]_{(0,0)+2(0,1)+2(1,0)+(1,1)}
     +[2,2,0]_{(\frac{1}{2},\frac{1}{2})}
     +[3,0,1]_{(\frac{1}{2},\frac{1}{2})}
     +[4,0,0]_{(0,0)}
     $ \\ \hline $
     \dDelta+\frac{9}{2}$& $
     [0,0,1]_{2(0,\frac{1}{2})+(0,\frac{3}{2})+3(1,\frac{1}{2})+2(1,\frac{3}{2})+(2,\frac{3}{2})}
     +[0,0,3]_{(\frac{1}{2},0)+(\frac{1}{2},1)}
     +[0,1,1]_{3(\frac{1}{2},0)+4(\frac{1}{2},1)+(\frac{1}{2},2)+(\frac{3}{2},0)+2(\frac{3}{2},1)}
     $\\& $
     +[0,1,3]_{(0,\frac{1}{2})}
     +[0,2,1]_{2(0,\frac{1}{2})+(0,\frac{3}{2})+2(1,\frac{1}{2})}
     +[0,3,1]_{(\frac{1}{2},0)}
     +[1,0,0]_{2(\frac{1}{2},0)+3(\frac{1}{2},1)+(\frac{3}{2},0)+2(\frac{3}{2},1)+(\frac{3}{2},2)}
     $\\& $
     +[1,0,2]_{2(0,\frac{1}{2})+(0,\frac{3}{2})+2(1,\frac{1}{2})+(1,\frac{3}{2})}
     +[1,1,0]_{3(0,\frac{1}{2})+(0,\frac{3}{2})+4(1,\frac{1}{2})+2(1,\frac{3}{2})+(2,\frac{1}{2})}
     +[1,1,2]_{(\frac{1}{2},0)+(\frac{1}{2},1)}
     $\\& $
     +[1,2,0]_{2(\frac{1}{2},0)+2(\frac{1}{2},1)+(\frac{3}{2},0)}
     +[1,3,0]_{(0,\frac{1}{2})}
     +[2,0,1]_{2(\frac{1}{2},0)+2(\frac{1}{2},1)+(\frac{3}{2},0)+(\frac{3}{2},1)}
     +[2,1,1]_{(0,\frac{1}{2})+(1,\frac{1}{2})}
     $\\& $
     +[3,0,0]_{(0,\frac{1}{2})+(1,\frac{1}{2})}
     +[3,1,0]_{(\frac{1}{2},0)}
     $ \\ \hline $
     \dDelta+5$& $
     [0,0,0]_{(\frac{1}{2},\frac{1}{2})+(\frac{1}{2},\frac{3}{2})+(\frac{3}{2},\frac{1}{2})+(\frac{3}{2},\frac{3}{2})}
     +[0,0,2]_{2(0,0)+(0,1)+(0,2)+2(1,1)}
     +[0,1,0]_{3(0,1)+3(1,0)+2(1,1)+(1,2)+(2,1)}
     $\\& $
     +[0,1,2]_{2(\frac{1}{2},\frac{1}{2})+(\frac{1}{2},\frac{3}{2})}
     +[0,2,0]_{3(\frac{1}{2},\frac{1}{2})+(\frac{1}{2},\frac{3}{2})+(\frac{3}{2},\frac{1}{2})}
     +[0,2,2]_{(0,0)}
     +[0,3,0]_{(0,1)+(1,0)}
     $\\& $
     +[1,0,3]_{(0,1)}
     +[1,1,1]_{2(0,0)+2(0,1)+2(1,0)+2(1,1)}
     +[1,2,1]_{(\frac{1}{2},\frac{1}{2})}
     +[2,0,0]_{2(0,0)+(1,0)+2(1,1)+(2,0)}
     $\\& $
     +[2,0,2]_{(\frac{1}{2},\frac{1}{2})}
     +[2,1,0]_{2(\frac{1}{2},\frac{1}{2})+(\frac{3}{2},\frac{1}{2})}
     +[2,2,0]_{(0,0)}
     +[3,0,1]_{(1,0)}
     +[1,0,1]_{4(\frac{1}{2},\frac{1}{2})+2(\frac{1}{2},\frac{3}{2})+2(\frac{3}{2},\frac{1}{2})+(\frac{3}{2},\frac{3}{2})}
     $ \\ \hline $
     \dDelta+\frac{11}{2}$& $
     [0,0,1]_{(\frac{1}{2},0)+2(\frac{1}{2},1)+(\frac{1}{2},2)+(\frac{3}{2},0)+(\frac{3}{2},1)}
     +[0,0,3]_{(0,\frac{3}{2})}
     +[0,1,1]_{3(0,\frac{1}{2})+(0,\frac{3}{2})+2(1,\frac{1}{2})+(1,\frac{3}{2})}
     +[0,2,1]_{(\frac{1}{2},0)+(\frac{1}{2},1)}
     $\\& $
     +[1,0,0]_{(0,\frac{1}{2})+(0,\frac{3}{2})+2(1,\frac{1}{2})+(1,\frac{3}{2})+(2,\frac{1}{2})}
     +[1,0,2]_{(\frac{1}{2},0)+2(\frac{1}{2},1)}
     +[1,1,0]_{3(\frac{1}{2},0)+2(\frac{1}{2},1)+(\frac{3}{2},0)+(\frac{3}{2},1)}
     $\\& $
     +[1,1,2]_{(0,\frac{1}{2})}
     +[1,2,0]_{(0,\frac{1}{2})+(1,\frac{1}{2})}
     +[2,0,1]_{(0,\frac{1}{2})+2(1,\frac{1}{2})}
     +[2,1,1]_{(\frac{1}{2},0)}
     +[3,0,0]_{(\frac{3}{2},0)}
     $ \\ \hline $
     \dDelta+6$& $
     [0,0,0]_{(0,0)+(0,2)+(1,1)+(2,0)}
     +[0,0,2]_{(\frac{1}{2},\frac{1}{2})+(\frac{1}{2},\frac{3}{2})}
     +[0,1,0]_{2(\frac{1}{2},\frac{1}{2})+(\frac{1}{2},\frac{3}{2})+(\frac{3}{2},\frac{1}{2})}    +[2,0,2]_{(0,0)}
     +[2,1,0]_{(1,0)}
     $\\& $
     +[0,1,2]_{(0,1)}
     +[0,2,0]_{2(0,0)+(1,1)}
     +[1,0,1]_{(0,0)+2(0,1)+2(1,0)+(1,1)}
     +[1,1,1]_{2(\frac{1}{2},\frac{1}{2})}
     +[2,0,0]_{(\frac{1}{2},\frac{1}{2})+(\frac{3}{2},\frac{1}{2})}
     $ \\ \hline $
     \dDelta+\frac{13}{2}$& $
     [0,0,1]_{(0,\frac{1}{2})+(0,\frac{3}{2})+(1,\frac{1}{2})}
     +[0,1,1]_{(\frac{1}{2},0)+(\frac{1}{2},1)}
     +[1,0,0]_{(\frac{1}{2},0)+(\frac{1}{2},1)+(\frac{3}{2},0)}
     +[1,0,2]_{(0,\frac{1}{2})}
     $\\& $
     +[1,1,0]_{(0,\frac{1}{2})+(1,\frac{1}{2})}
     +[2,0,1]_{(\frac{1}{2},0)}
     $ \\ \hline $
     \dDelta+7$& $
     [0,0,0]_{(\frac{1}{2},\frac{1}{2})}
     +[0,0,2]_{(0,0)}
     +[0,1,0]_{(0,1)+(1,0)}
     +[1,0,1]_{(\frac{1}{2},\frac{1}{2})}
     +[2,0,0]_{(0,0)}
     $ \\ \hline $
     \dDelta+\frac{15}{2}$ & $
     [0,0,1]_{(\frac{1}{2},0)}
     +[1,0,0]_{(0,\frac{1}{2})}
     $ \\ \hline $
     \dDelta+8$& $
     [0,0,0]_{(0,0)}
     $ \\ \hline
     \end{tabular}}
 \caption{Long Konishi multiplet \label{Ktable}}
\end{table}


\bigskip

\end{document}